\documentclass[aps,pra,twocolumn,groupedaddress,showpacs,showkeys,preprintnumbers]{revtex4}
\usepackage{graphicx}
\usepackage{amsmath}
\usepackage{amsfonts}

\begin{document}

\preprint{D0267}

\title{Charge of a particle generated by a captured pair of photons}


\author{Herbert Wei\ss}
\email[\emph{Email: }]{hw@terminal-e.de}
\affiliation{Am Bucklberg 19, D-83620 Feldkirchen-Westerham, Germany}

\date{\today}

\begin{abstract}
The model of charge generation is based on the wave model of a particle, for which a brief description is given. The particle is comprised of two photons, captured in the volume of the particle, and reveals the complete relativistic behaviour. The electromagnetic waves of the two photons are supposed to induce an electrostatic field. The surface integral of this electrostatic field is identified with the charge of the particle. The magnitude of the electric field is derived from the energy density of the photons. The volume and the surface are related to the wavelength of the photons, thus leading to a charge of the particle, which is independent of its mass. 

\end{abstract}


\maketitle




\section{Introduction}

In classical mechanics, as well as in quantum mechanics, the positive or negative charge of a particle is postulated to be an integer multiple of elementary charge \(e\), with a numerical value of \mbox{1.602\:176\:53 \(\times 10^{-19}\) Coulomb (As)} in SI units. Charge is a physical quantity with outstanding properties: In nature, it can achieve only two values, \(+e\) and \(-e\). The charge of a particle is independent from its mass. Charge remains constant under any coordinate transformation, the law of charge conservation is independent of the metric used, it is a premetric law \cite{Hehl:physics/0407022v1}. -- Nevertheless, there is only very few knowledge about the nature of charge.

Naudts and Kuna \cite{Naudts:hep-th/0012209v1} give a new formulation of the particle model from Doplicher, Fredenhagen, and Roberts, yielding the expression \(b^2 = \lambda h c_0 / \left(2 \pi \right)\) for the charge \(b\) of the proton. \(\lambda\) is arbitrarily chosen and is assumed to be the fine structure constant \(\alpha\). -- Chernitskii \cite{Chernitskii:hep-th/0002083v1} takes the idea of dyons from Schwinger. Dyons are entities which carry electric charge as well as magnetic charge. He creates bidyons in taking two dyons a distance \(2a\) apart with equally sized electric charges \(d\) of same polarity and equally sized magnetic charges \(b\) of opposite polarity in order to cancel the net magnetic flux of a closed surface covering the particle. Dyons are supposed to have either electric charge and/or magnetic charge. -- Azc\`arraga et. al. \cite{Azcarraga:hep-th/0510161v1} report from a relativistic particle model using a pair of twistors in a Liouville one-form which leads to an enhanced phase space of 18 dimensions. Quantization provides a set of wave equations determining mass, spin and electric charge of a relativistic particle. The sum of internal scalar charge \(t_0\) and \(t_3\) equals the electric charge \(e_p\) of the particle. Subsequently, constraint \(R_6\) equates \(e_p\) with elementary charge \(e\). -- Fedoruk et. al. \cite{Fedoruk:hep-th/0510266v1} extend the Shirafuji model using a two-twistor description and quantize the model in order to obtain wavefunctions describing relativitstic particles with mass, spin, and electric charge. They predict numerical values coming from the solution of suitable chosen constraints. The symbol \(q\) enters constraint 4.2d for the charge of the particle, and \(q\) is identified with elementary charge \(e\). -- A merely speculative attempt to give an explanation of charge, independent from \(e\), comes from Baten \cite{Baten:physics/0009085v1}, who suggests to relate electric charge \(+e\) and \(-e\) to possible squeeze phases \(+1\) or \(-1\) of a hypothetical electromagnetic protofield movement towards the so-called reduction center of a particle. -- Another approach is that of Hadley \cite{Hadley:physics/0601032v1}, who states that Stokes theorem has limited application in manifolds that are not time orientable. He says:
``when the theorems do not apply, it is possible to have the appearance of charge arising from the source free equations, because there can be a net surface flux with zero enclosed charge.''
The non-zero net flux can give rise to either a virtual electric charge as well as a magnetic charge. -- One can imagine the difficulties and the enormous effort arising from the theoretical framework necessary to provide a solid basis for a theory violating Stokes' or Gaussian theorem in order to yield an expression for the electric charge of a particle. We mention this, because our approach positively applies Gaussian theorem in the usual manner, and it is quite sufficient to yield a mechanism for creating charge without endeavoring a special particle carrying elementary charge \(e\), as will be shown below.

The models of charge, given in \cite{Naudts:hep-th/0012209v1, Chernitskii:hep-th/0002083v1, Azcarraga:hep-th/0510161v1, Fedoruk:hep-th/0510266v1}, have reference to elementary charge \(e\), meaning that the models are adjusted to comply with observations. The charge model presented here suffers from the same problem, it requires the constant \(C_W\). Future development of the model is supposed to resolve constant \(C_W\).

The mechanism of charge generation is based on the wave model of a particle given in  \cite{Weiss:Wellenmodell}. In order to provide a sufficient basis for the charge model, we give a brief description of the wave model in section \ref{SEC:WAVEMODEL}. The actual model of charge generation is presented in section \ref{SEC:CHARGE}. -- All equations are given exclusively in SI units.

\section{Wave model of a particle\label{SEC:WAVEMODEL}}

Since Max Planck postulated in 1900 the electromagnetic radiation being emitted and absorbed by discret quanta \cite{Planck:Quantum} and the correct interpretation of the photo effect in 1905 by Albert Einstein \cite{Einstein:Photoeffect}, the dual nature of electromagnetic waves was evident. Photons possess properties of waves according to Maxwell equations as well as properties of particles with momentum \(p\) according to Eq.\ (\ref{Q:MOMENTUM}),
\begin{equation}
	p = \frac{h}{\lambda} , \label{Q:MOMENTUM}
\end{equation}

with \(h\) the Planck's constant and \(\lambda\) the wavelength of the photon. De Broglie's idea was to assume a similar duality for particles \cite{deBroglie:1923}. He relatet the momentum \(p\) of a particle with mass \(m\) and speed \(v\), Eq.\ (\ref{P:MOMENTUM}), to a matter wave of wavelength \(\lambda\), Eq.\ (\ref{P:LAMBDA}).
\begin{eqnarray}
	p &=& mv \label{P:MOMENTUM} \\ \nonumber \\
	\lambda &=&  \frac{h}{mv} \label{P:LAMBDA}
\end{eqnarray}

Four years later, in 1927, Davisson and Germer \cite{Davisson:Germer:1927} experimentally confirmed de Broglie's postulate by a diffraction experiment with an electron beam directed on Ni-crystals. The year before Erwin Schr\"odinger used the idea of matter waves and defined a wave equation for particles, the Schr\"odinger equation \cite{Schroedinger:1926}. The solution of this wave equation gave the first reasonable explanation for the discrete states of an electron in an atom and was the start of quantum mechanics.
The dual properties of particles are commonly accepted. The wave properties are derived from equality (\ref{Q:MOMENTUM}), which is originally valid for light quanta. For this reason the conjecture is near at hand that a particle \emph{consists} of light quanta. This idea was used in the wave model of a particle.

\subsection{Definition of wave model}

\begin{figure}[h]
	\centering
		\includegraphics{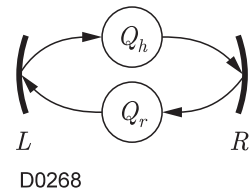}
	\caption{Wave model of a particle with quanta \(Q_h\) and \(Q_r\). The wave of \(Q_h\) is reflected at the right-hand wall \(R\) and the reflected part contributes to quantum \(Q_r\). The wave of \(Q_r\) is reflected at the left-hand wall \(L\) and the reflected part contributes to quantum \(Q_h\).}
	\label{fig:d0268}
\end{figure}

A particle \(P\) is assumed to consist of two light quanta \(Q_h\) and \(Q_r\) traveling in opposite directions (Fig.\ \ref{fig:d0268}).
\begin{equation}
	P = \{ Q_h, Q_r \}
	\label{PARTICLE}
\end{equation}

For convenience we assume these directions being aligned with the \(x\)-axis. The light quanta are captured between two walls \(L\) and \(R\). We do not specify the nature of the walls, we treat them being able to reflect the electromagnetic wave of the captured photons, regardless of the energy. Quantum \(Q_h\) travels in the positive \(x\)-direction, is reflected by the right-hand wall \(R\), and the reflected part contributes to quantum \(Q_r\). Quantum \(Q_r\) travels in the negative \(x\)-direction, is reflected by the left-hand wall \(L\), and the reflected part contributes to quantum \(Q_h\). The two walls must be held at a certain distance \(d_x\) in order to enable contructive interference of the two waves.

Tables \ref{TAB:PARTICLE} and \ref{TAB:QH:QR} list the quantities pertaining to the model. All vectorial quantities are represented by their \(x\)-component with the corresponding basis vector \(\widehat{\bf x}\) being omitted, the \(y\)- and \(z\)-components are assumed to be zero unless otherwise noticed.
\def\TTAB{\hspace*{53 mm}}
\begin{table}[h]
	\centering
	\caption{Quantities assigned to the particle \(P\)\label{TAB:PARTICLE}}
		\begin{ruledtabular}
		\begin{tabular}{c l}
			symbol 	& description \TTAB \\ \hline
			\(Q_h\) &	quantum traveling in the positive \(x\)-direction \\
			\(Q_r\) &	quantum traveling in the negative \(x\)-direction \\
			\(R\)   &	right-hand wall \\
			\(L\)   &	left-hand wall \\
			\(v\)   &	drift velocity in the positive \(x\)-direction\\
			\(W_g\) &	energy of the particle \\
			\(p_g\) &	momentum of the particle in positive \(x\)-direction \\
			\(n_g\) &	wave count of the particle  \\
			\(\lambda_d\) &	de-Broglie wavelength of the particle \\
			\(d_x\) & dimension parallel to the \(x\)-axis \\
			\(d_\perp\) & dimension perpendicular to the \(x\)-axis\\
			\(s_g\) & propagation length of a complete cycle\\
		\end{tabular}
		\end{ruledtabular}
\end{table}

\begin{table}[h]
	\centering
	\caption{Quantities assigned to quanta \(Q_h\) and \(Q_r\)\label{TAB:QH:QR}}
		\begin{ruledtabular}
		\begin{tabular}{c l}
			symbol 	& description \TTAB \\ \hline
			\(f_h\) & frequency of \(Q_h\) \\
			\(\lambda_h\) & wavelength of \(Q_h\) \\
			\(n_h\) & wave count of \(Q_h\) \\
			\(W_h\) & energy of \(Q_h\) \\
			\(p_h\) & momentum of \(Q_h\) \\
			\(s_h\) & propagation length of  \(Q_h\)\\  \hline
			\(f_r\) & frequency of \(Q_r\) \\
			\(\lambda_r\) & wavelength of \(Q_r\) \\
			\(n_r\) & wave count of \(Q_r\) \\
			\(W_r\) & energy of \(Q_r\) \\
			\(p_r\) & momentum of \(Q_r\) \\
			\(s_r\) & propagation length of  \(Q_r\)
		\end{tabular}
		\end{ruledtabular}
\end{table}

The energy \(W_g\) of the particle is simply the sum of energies \(W_h\) and \(W_r\) of the two light quanta
\begin{equation}
	W_g = W_h + W_r ,
	\label{DEF:ENERGY}
\end{equation}

and the energies \(W_h\) and \(W_r\) are given by formulae
\begin{eqnarray}
	W_h & = & h f_h \label{DEF:EH}
	\hspace*{20 mm} \\ {\rm and} \hspace{6 mm}
	W_r & = & h f_r \label{DEF:ER} ,
\end{eqnarray}

with \(h\) denoting Planck's constant. Putting Eqs.\ (\ref{DEF:EH}) and (\ref{DEF:ER}) into Eq.\ (\ref{DEF:ENERGY}) yields Eq.\ (\ref{DEF:EFHFR}) defining the energy \(E_g\) of the particle in terms of the frequencies \(f_h\) and \(f_r\).
\begin{equation}
	W_g = h \left(f_h + f_r\right)
	\label{DEF:EFHFR}
\end{equation}

The momentum \(p_g\) of the particle is the sum of momenta \(p_h\) and \(p_r\)
\begin{equation}
	p_g = p_h + p_r ,
	\label{DEF:MOMENTUM}
\end{equation}

with \(p_h\) and \(p_r\) given by formula (\ref{DEF:PH}) and (\ref{DEF:PR}).
\begin{eqnarray}
	p_h & = & \frac{h}{c_0} f_h \label{DEF:PH}	\\ 
	p_r & = & -\frac{h}{c_0} f_r \label{DEF:PR}
\end{eqnarray}

Momentum \(p_r\) must be less than zero, because \(Q_r\) is traveling in the negative \(x\)-direction. Again, putting Eqs.\ (\ref{DEF:PH}) and (\ref{DEF:PR}) into Eq.\ (\ref{DEF:MOMENTUM}) yields Eq.\ (\ref{DEF:PFHFR}) defining the momentum \(p_g\) of the particle in terms of the frequencies \(f_h\) and \(f_r\).
\begin{equation}
	p_g = \frac{h}{c_0} \left(f_h - f_r\right)
	\label{DEF:PFHFR}
\end{equation}

The wavelengths \(\lambda_h\) and \(\lambda_r\) of the particle are simply:
\begin{eqnarray}
	\lambda_h & = & \frac{c_0}{f_h} \label{DEF:LH}
	\hspace*{20 mm} \\ {\rm and} \hspace{6 mm}
	\lambda_r & = & \frac{c_0}{f_r} \label{DEF:LR} .
\end{eqnarray}

In order to obtain constructive interference, the sum \(n_g\) of wavecounts \(n_h\) and \(n_r\) must be an integer,
\begin{equation}
	n_g = n_h + n_r ,   	\label{DEF:WAVECOUNT}
\end{equation}

with conditions (\ref{CON:NH:NR}) and (\ref{CON:NP}).
\begin{eqnarray}
	n_h, n_r & \in & {\mathbb R} \label{CON:NH:NR} \\
	     n_g & \in & {\mathbb N} \label{CON:NP}
\end{eqnarray}

Length \(d_x\) of the particle in the \(x\)-direction equals the product of wave count \(n_h\) and wavelength \(\lambda_h\), as well as the product of wave count \(n_r\) and wavelength \(\lambda_r\).
\begin{equation}
	d_x = n_h \lambda_h = n_r \lambda_r \label{DIM:X}
\end{equation}

From Eq.\ (\ref{DIM:X}) we get a relation between wave counts and wavelengths.
\begin{equation}
	\frac{n_h}{n_r} = \frac{\lambda_r}{\lambda_h} \label{REL:WAVECOUNT:LAMBDA}
\end{equation}

Using Eqs.\ (\ref{DEF:LH}) and (\ref{DEF:LR}) Eq.\ (\ref{REL:WAVECOUNT:LAMBDA}) aquires the form (\ref{REL:WAVECOUNT:FREQ}):
\begin{equation}
	\frac{n_h}{n_r} = \frac{f_h}{f_r} \label{REL:WAVECOUNT:FREQ}
\end{equation}

The propagation length \(s_g\) is the sum of distances \(s_h\) and \(s_r\) a state of the electromagnetic wave propagates, starting at the left-hand wall \(L\), traveling to the right-hand wall \(R\), and traveling back to the left-hand wall \(L\) after reflection at the right-hand wall \(R\). Distances \(s_h\) and \(s_r\) are measured in the frame at rest, even for the moving particle.
\begin{equation}
	s_g = s_h + s_r \label{DEF:SG}
\end{equation}

We define the intrinsic period \(T\) of the particle, measured in the system at rest, as the time elapsed required for a state of the electromagnetic wave to travel the distance \(s_g\).
\begin{equation}
	T = \frac{s_g}{c_0} \label{DEF:T}
\end{equation}

\subsection{Particle at rest\label{SSEC:PATREST}}

At rest, the frequencies \(f_h\) and \(f_r\) of both quanta \(Q_h\) and \(Q_r\) are equally and have magnitude \(f_0\).
\begin{equation}
	v=0 \hspace{6 mm} \Rightarrow \hspace{6 mm} f_h(0)=f_r(0)=f_0
	\label{EQ:F0}
\end{equation}

From Eqs.\ (\ref{DEF:EFHFR}) and (\ref{DEF:PFHFR}) we obtain the energy \(W_g(0)\) and the momentum \(p_g(0)\) of the particle at rest.
\begin{eqnarray}
	W_g(0) & = & h f_0 \label{EQ:EN0} \\
	p_g(0) & = & \frac{h}{c_0} \left(f_0 - f_0\right) \equiv 0 \label{EQ:PN0}
\end{eqnarray}

From Eqs.\ (\ref{EQ:F0}), (\ref{DEF:LH}), and (\ref{DEF:LR}) we get wavelengths \(\lambda_h(0)\) and \(\lambda_r(0)\) by Eq.\ (\ref{DEF:LH0:LR0}). 
\begin{equation}
	\lambda_h(0) =\lambda_r(0)= \lambda_0 = \frac{c_0}{f_0} \label{DEF:LH0:LR0}
\end{equation}

The quanta at rest have equally wave lengths and hence equally wave counts due to Eq.\ (\ref{REL:WAVECOUNT:LAMBDA}).
\begin{equation}
	n_h(0) =n_r(0)= n_0 = \frac{n_g}{2} \label{EQ:NH0:NR0}
\end{equation}

Lentgh \(d_x(0)\) of the particle at rest is given by formula (\ref{DX0}).
\begin{equation}
	d_x(0) = n_0 \lambda_0 = \frac{n_g c_0}{2 f_0}
	\label{DX0}
\end{equation}

In the frame at rest, isotropy of speed of light is assumed. Regarding the propagation of an electromagnetic wave in directions different from \(x\), the particle at rest must have the shape of a sphere in order to maintain constructive interference in these directions. Thus, the diameter of the particle in any direction equals \(d_x(0)\), which we denote as \(d_0\).
\begin{equation}
	d_0 = \frac{n_g c_0}{2 f_0} \label{DEF:D0} 
\end{equation}

Then, the surface of a particle resting in the origin of the coordinate sytem is given by Eq.\ (\ref{SURFACE:AT:REST}),
\begin{equation}
	x^2 + y^2 + z^2 = \frac{d_0^2}{4} ,
	\label{SURFACE:AT:REST}
\end{equation}

and the diameters \(d_y\) and \(d_z\), perpendicular to direction \(x\), are given by Eq.\ (\ref{EQ:DY0:DZ0}).
\begin{equation}
	d_y(0) = d_z(0) = d_{\perp} = d_0  \label{EQ:DY0:DZ0} 
\end{equation}

Distances \(s_h\) and \(s_r\) are equally and are the same as length \(d_0\), so that the sum \(s_g\) is given by Eq.\ (\ref{EQ:SHV:0})
\begin{equation}
  s_g = 2 d_0 = \frac{n_g c_0}{f_0}\label{EQ:SHV:0}
\end{equation}

The intrinsic time period \(T_0\) of the particle at rest is given by Eq.\ (\ref{T0}).
\begin{equation}
	T_0 = \frac{s_g}{c_0} = \frac{n_g}{f_0} \label{T0}
\end{equation}

\subsection{Particle moving at velocity \(v\)}

Now, the particle moves at velocity \(v\) in the positive \(x\)-direction. That means, the two walls are moving at speed \(v\). The frequencies of quanta \(Q_h\) and \(Q_r\) must be different because of the Doppler effect. Considering the reflection of the wave at the right-hand wall \(R\) yields relation (\ref{REL:FH:FR}) of frequency \(f_h\) to frequency \(f_r\).
\begin{equation}
	\frac{f_h}{f_r} = \frac{c_0 + v}{c_0 - v}
	\label{REL:FH:FR}
\end{equation}

Here, we encounter the problem that the relation to the particle at rest is lost. The identity of the particle at rest is the frequency \(f_0\). In order to recover the relation, we introduce condition (\ref{DEF:F0:FH:FR}): the geometrical mean of frequencies \(f_h\) and \(f_r\) must be equal to the frequency \(f_0\).
\begin{equation}
	\boxed{ \sqrt{f_h f_r} = f_0 }
	\label{DEF:F0:FH:FR}
\end{equation}

With Eq.\ (\ref{DEF:F0:FH:FR}) the frequencies \(f_h\) and \(f_r\) of the moving particle can be calculated from Eq.\ (\ref{REL:FH:FR}) and yields
\begin{eqnarray}
	f_h & = & f_0 \sqrt{\frac{c_0 + v}{c_0 - v}} \label{EQ:FH:V}
	\hspace*{20 mm} \\ {\rm and} \hspace*{10 mm}
	f_r & = & f_0 \sqrt{\frac{c_0 - v}{c_0 + v}} \label{EQ:FR:V} .
\end{eqnarray}

Using formula (\ref{EQ:FSUM}) and (\ref{EQ:FDIF}) for the sum and the difference of frequencies \(f_h\) and \(f_r\),
\begin{eqnarray}
	f_s & = & f_h + f_r = \gamma \: 2 f_0, \label{EQ:FSUM}
	\hspace*{20 mm} \\ {\rm and} \hspace*{10 mm}
	f_d & = & f_h - f_r = \gamma \: 2 f_0\frac{v}{c_0}, \label{EQ:FDIF} 
\end{eqnarray}

we obtain energy \(W_g\) and momentum \(p_g\) of the moving particle from Eqs.\ (\ref{DEF:EFHFR}) and (\ref{DEF:PFHFR}),
\begin{eqnarray}
	W_g & = & \gamma \: 2 h f_0  \label{EQ:EG:V}, \\
	p_g & = & \gamma \: \frac{2 h f_0 v}{c_0^2}, \label{EQ:PG:V}
\end{eqnarray}
 
with the gamma factor given by (\ref{GAMMA}).
\begin{equation}
	\gamma = \frac{1}{\sqrt{1-\frac{v^2}{c_0^2}}} \label{GAMMA}
\end{equation}

Using Eqs.\ (\ref{DEF:WAVECOUNT}) and (\ref{REL:WAVECOUNT:FREQ}) the wave counts \(n_h\) and \(n_r\) can be calculated from Eqs.\ (\ref{EQ:FH:V}) and (\ref{EQ:FR:V}).
\begin{eqnarray}
	n_h = \frac{n_g}{2}\left( 1 + \frac{v}{c_0}\right) \label{EQ:NH:V} \\
	n_r = \frac{n_g}{2}\left( 1 - \frac{v}{c_0}\right) \label{EQ:NR:V} 
\end{eqnarray}

The length \(d_x\) of the moving particle is from Eqs.\ (\ref{DEF:LH}), (\ref{DIM:X}), (\ref{EQ:FH:V}), and (\ref{EQ:NH:V})
\begin{equation}
	d_x = d_0 \sqrt{1-\frac{v^2}{c_0^2}} .
	\label{DX:V}
\end{equation}

The propagation length \(s_g\) of the moving particle is given by Eq.\ (\ref{EQ:SG:V}).
\begin{equation}
	s_g = \gamma \frac{n_g c_0}{f_0}  \label{EQ:SG:V}
\end{equation}

The intrinsic time period \(T\) of the moving particle is given by Eq.\ (\ref{EQ:TI:V}).
\begin{equation}
	T = \gamma \frac{n_g}{f_0}  \label{EQ:TI:V}
\end{equation}

Considering the propagation of the electromagnetic wave in directions different from \(x\) and taking into account constructive interference, the moving particle must aquire the form of a rotational ellipsoid according to Eq.\ (\ref{SURFACE:AT:V}).
\begin{equation}
	\gamma^2 x^2 + y^2 + z^2 = \frac{d_0^2}{4} ,
	\label{SURFACE:AT:V}
\end{equation}

\subsection{Properties of the wave model}

The wave model of a particle is derived from postulate (\ref{PARTICLE}) and condition (\ref{DEF:F0:FH:FR}) and yields the complete relativistic behaviour of matter:
\begin{itemize}
	\item {\scshape Length contraction} is demonstrated by Eqs.\ (\ref{SURFACE:AT:REST}) and (\ref{SURFACE:AT:V}) 
	\item {\scshape Time dilation} is demonstrated by Eqs.\ (\ref{T0}) and (\ref{EQ:TI:V})
	\item {\scshape Relativistic energy} is demonstrated by Eqs.\ (\ref{EQ:EN0}) and (\ref{EQ:EG:V})
	\item {\scshape Relativistic momentum} is demonstrated by Eqs.\ (\ref{EQ:PN0}) and (\ref{EQ:PG:V})
\end{itemize}

Moreover, the wave model is able to give an explanation for the enhanced lifetime of moving particles. The model predicts a particle to posses an internal clock, which is comprised of the captured photons bouncing between the two walls. The time elapsed between two ticks becomes longer the faster the particle moves.

\subsection{de Broglie wavelength \(\lambda_d\) and phase speed \(c_\varphi\)}

Equation (\ref{DEF:PFHFR}) relates the momentum of the particle to the difference \(f_d\) of frequencies \(f_h\) and \(f_r\).
\begin{equation}
	f_d = f_h - f_r
	\label{DEF:DIFF}
\end{equation}

Although the difference frequency \(f_d\) has no real representation, for instance, by a photon of energy \(h f_d\) captured within the walls, we can assign the wavelength \(\lambda_d\) according to Eq.\ (\ref{EQ:LAMBDA:D}).
\begin{equation}
	\lambda_d = \frac{c_0}{f_d}
	\label{EQ:LAMBDA:D}
\end{equation}

When we replace the difference frequency \(f_d\) with the right-hand side of Eq.\ (\ref{EQ:FDIF})
\begin{equation}
	\lambda_d = \frac{c_0}{\gamma \: 2 f_0 \frac{v}{c_0}} \label{EQ:LAMBDA:D2}
\end{equation}

and enhance numerator and denominator with Planck's constant \(h\), Eq.\ (\ref{EQ:LAMBDA:D2}) aquires the form (\ref{EQ:LAMBDA:D3}) after some simple rearrangement.
\begin{equation}
	\lambda_d = \frac{h}{\frac{\gamma \: 2 h f_0}{c_0^2}v} =\frac{h}{\frac{W_g}{c_0^2}v} \label{EQ:LAMBDA:D3}
\end{equation}

Using Einstein's energy mass equivalence formula,	\(W_g = m_g c_0^2\), we obtain finally Eq.\ (\ref{EQ:LAMBDA:DB}) 
\begin{equation}
	\lambda_d = \frac{h}{m_gv}, \label{EQ:LAMBDA:DB}
\end{equation}

with \(m_g\) the correct relativistic mass of the moving particle. Equation (\ref{EQ:LAMBDA:DB}) is identical with Eq.\ (\ref{P:LAMBDA}), meaning that the wave model of a particle yields an explanation for matter waves and for the de Broglie wavelength.

When the de Broglie wavelength is used to calculate the total energy of the particle employing Planck's formula \(E=h f\), the corresponding frequency \(f_B\) must be calculated from a propagation speed other than speed of light. The correct frequency is obtained using a propagation speed of \(c_\varphi = c_0^2/v\). This is called the \emph{phase speed} in order to avoid defining a propagation speed higher than speed of light.
\begin{eqnarray}
	f_B &=& \frac{c_\varphi}{\lambda_d} \label{EQ:FD:VPHI} \\
	c_\varphi &=& \frac{c_0^2}{v} \label{EQ:VPHI}
\end{eqnarray}

There is no mechanism known which is able to expound such an odd behaviour of this matter wave, especially the phenomenon that with velocity \(v = 0\) the phase speed becomes infinity. Now, this problem can be solved with the wave model. At time \(t_0\) we take a point on the \(x\)-axis where the phases \(\varphi_h\) and \(\varphi_r\) sum up to phase \(\varphi_s\),
\begin{equation}
	\varphi_h + \varphi_r = \varphi_s,	\label{EQ:SUM:PHI:ZERO}
\end{equation}

and look for the condition of constant phase sum \(\varphi_s\).
The phases \(\varphi_h\) and \(\varphi_r\) of the wave functions for \(Q_h\) and \(Q_r\), respectively, are given by Eqs.\ (\ref{EQ:PHI:H}) and (\ref{EQ:PHI:R}).
\begin{eqnarray}
	\varphi_h &=& 2 \pi f_h \left( t - \frac{x}{c_0} \right) \label{EQ:PHI:H} \\
	\varphi_r &=& 2 \pi f_r \left( t + \frac{x}{c_0} \right) \label{EQ:PHI:R}
\end{eqnarray}

We substitute \(\varphi_h\) and \(\varphi_r\) in Eq.\ (\ref{EQ:SUM:PHI:ZERO}) with the right-hand side of Eqs.\ (\ref{EQ:PHI:H}) and (\ref{EQ:PHI:R}).
\begin{eqnarray}
	2 \pi f_h \left( t - \frac{x}{c_0} \right)
	 + 2 \pi f_r \left( t + \frac{x}{c_0} \right) &=& \varphi_s	\nonumber \\
	2 \pi \left(f_h + f_r\right) t - 2 \pi \left( f_h - f_r \right)\frac{x}{c_0}
	 &=& \varphi_s \nonumber \\
			2 \pi f_s t - 2 \pi f_d \frac{x}{c_0} &=& \varphi_s
	\label{EQ:SUM2:PHI:ZERO}
\end{eqnarray}

Using Eqs.\ (\ref{EQ:FSUM}) and (\ref{EQ:FDIF}) for the sum \(f_s\) and the difference \(f_d\) of frequencies \(f_h\) and \(f_r\), we obtain:
\begin{eqnarray}
			 t - \frac{v}{c_0^2} x
			 	 &=& \frac{\varphi_s}{2 \pi \gamma \: 2 f_0} \nonumber \\
			  x(\varphi_s, t)
			   &=& \frac{c_0^2}{v} t - \frac{c_0^2 \varphi_s}{v 2 \pi \gamma \: 2 f_0}
	\label{EQ:SUM4:PHI:ZERO}
\end{eqnarray}

This is the equation for the location \(x(\varphi_s, t)\) as a function of constant sum \(\varphi_s\) of phases \(\varphi_h\) and \(\varphi_r\) and time \(t\). The time derivative of \(x\) is the speed the state of constant phase \(\varphi_s\) propagates,
\begin{equation}
  \frac{{\rm d} }{{\rm d} t}x(\varphi_s, t) = c_\varphi = \frac{c_0^2}{v}, \label{EQ:PHI:SPEED}
\end{equation}

which is identified with the phase speed \(c_\varphi\).

\section{Charge of the particle\label{SEC:CHARGE}}

Maxwell equation of divergence (\ref{MAX:DIV:E}) gives a relation between the electric field \({\bf E}_q\) and the charge density \(\rho\).
\begin{equation}
	\nabla {\bf E}_q = \frac{\rho}{\varepsilon_0}
	\label{MAX:DIV:E}
\end{equation}

We read this equation from the right-hand side to the left-hand side, i.e., we suppose the charge vector field \({\bf E}_q\) makes the particle appear to posses charge \(q\).
Using Gaussian law the charge \(q\) is given by the surface integral of vector field \({\bf E}_q\) over the surface \(A\) enclosing the particle.
\begin{equation}
	q = \iint \limits_{A} \varepsilon_0 \: {\bf E}_q \: {\rm d}{\bf A}
	\label{GAUSS}
\end{equation}

\({\rm d}{\bf A}\) is a vector of length d\(A\) pointing outside the surface of the particle.

\subsection{Charge \(q\) in terms of \(E_w\)}

\begin{figure}
	\centering
		\includegraphics{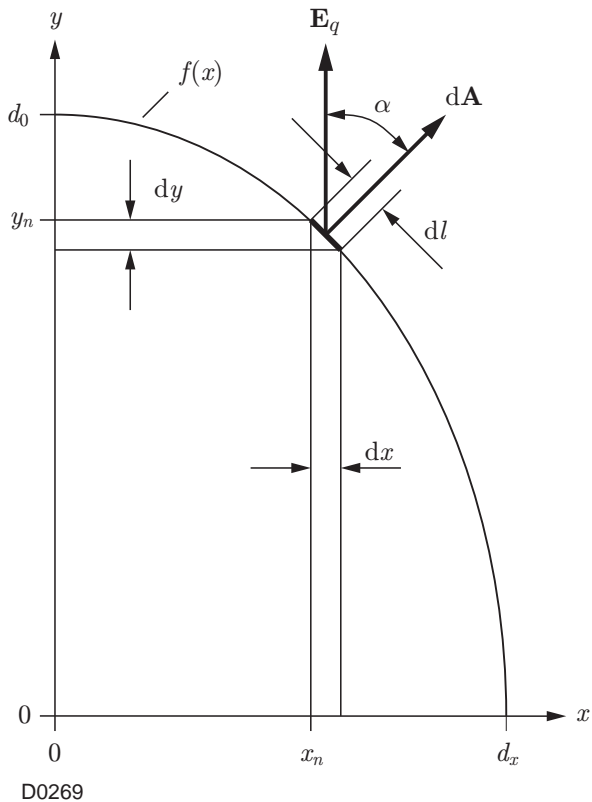}
	\caption{Contour of the particle, cut in the \(x\)-\(y\)-plane. Location of vector \({\bf E}_q\) relative to surface element d{\bf A}. The bold emphasized line element d\(l\) contributes to the surface element d{\bf A}. Line \(f(x)\) is the outline of the particle.}
	\label{fig:d0269}
\end{figure}

We assume the field vector \({\bf E}_q\) has a magnitude \(E_q\) proportional to the magnitude \(E_w\) of field vector \({\bf E}_w\) at the surface of the particle. \({\bf E}_w\) is the field vector of the electromagnetic wave pertaining to the photons \(Q_h\) and \(Q_r\). Vector \({\bf E}_q\) is common to the \(x\)-\({\rm d}{\bf A}\)-plane but perpendicular to the \(x\)-axis, thus having angle \(\alpha\) with vector \({\rm d}{\bf A}\), Fig.\ \ref{fig:d0269}.
\begin{equation}
	\alpha = \angle \: \left( {\bf E}_q, {\rm d}{\bf A} \right)	\label{EQ:ALPHA}
\end{equation}

Constant \(C_E\) denotes the ratio of \(E_q\) and \(E_w\).
\begin{equation}
	E_q = C_E E_w	\label{EQ:EQ:EW}
\end{equation}

With the relations (\ref{GAUSS}) and (\ref{EQ:EQ:EW}) charge \(q\) is given by
\begin{equation}
	q = \iint \limits_{A} \varepsilon_0 \: C_E \: E_w \: {\rm d}A \cos \alpha
	\label{EQ:GAUSS:2}
\end{equation}

The particle has rotational symmetry with respect to the \(x\)-axis, thus the contour of the cross section is an ellipsis with diameter \(d_x\) in the \(x\)-direction and diameter \(d_0\) perpendicular to \(x\). The function \(f(x)\) of the contour is
\begin{equation}
	f(x) = \frac{d_0}{d_x} \sqrt{\frac{d_x^2}{4}-x^2}  \label{EQ:F:VON:X}
\end{equation}

Surface element \({\rm d}A\) equals the contour line element d\(l\) times \(f(x)\) times angle d\(\varphi\).
\begin{equation}
	{\rm d}A = {\rm d}l \: f(x) \: {\rm d}\varphi ,  \label{EQ:DA}
\end{equation}

with the line element d\(l\) given by
\begin{equation}
	\left({\rm d}l\right)^2 = \left({\rm d}x\right)^2 + \left[f'(x)\:{\rm d}x\right]^2 , \label{EQ:DL}
\end{equation}

and \(\cos \alpha\) the ratio of d\(x\) and d\(l\).
\begin{equation}
	\cos \alpha = \frac{{\rm d}x}{{\rm d}l}		\label{EQ:COS:ALPHA}
\end{equation}

We plug in Eqs.\ (\ref{EQ:F:VON:X}-\ref{EQ:COS:ALPHA}) into Eq.\ (\ref{EQ:GAUSS:2}) and introduce the limits of the integrals, 0 and \(2 \pi\) for d\(\varphi\), and \(-d_x/2\) and \(+d_x/2\) for d\(x\).
\begin{equation}
	q = \int \limits_{0}^{2 \pi} \int \limits_{-d_x/2}^{+d_x/2}
				\varepsilon_0 \: C_E \: E_w \: {\rm d}l \:
			 \frac{d_0}{d_x} \sqrt{\frac{d_x^2}{4}-x^2}
			 \: \frac{{\rm d}x}{{\rm d}l} \: {\rm d}\varphi 
	\label{EQ:GAUSS:3}
\end{equation}

The first integral over d\(\varphi\) can be immediately executed and yields \(2 \pi\), line element d\(l\) is cancelled, and the remaining integral,
\begin{equation}
	q = 2 \pi \: \varepsilon_0 \: C_E \: E_w \: \frac{d_0}{d_x} \:
			 \int \limits_{-d_x/2}^{+d_x/2}
			  \sqrt{\frac{d_x^2}{4}-x^2}
			 \: {\rm d}x ,
	\label{EQ:GAUSS:4}
\end{equation}

is solved using {\small \footnotesize
	\[ \int \limits_{x_1}^{x_2} \sqrt{a^2 - x^2}\:  {\rm d}x
	= \frac{1}{2} \: \left[ x \sqrt{a^2 - x^2} + a^2 \arcsin \frac{x}{a} \right]^{x_2}_{x_1}.
\]}

\begin{equation}
	q  =  \varepsilon_0 \: C_E \: E_w \: \frac{d_0}{2} \:
			  \frac{d_x}{2} \: \pi^2
			  	\label{EQ:GAUSS:5}
\end{equation}

\subsection{Magnitude \(E_w\) in terms of energy \(W_g\)}

In order to determine the magnitude of vector \({\bf E}_w\), we equate the energy \(W_w\) of an electromagnetic wave with the energy \(W_g\) of a particle.
\begin{equation}
	W_w = W_g
	\label{EQ:WW:WG}
\end{equation}

The energy \(W_w\) is equal to the energy density \(w\) times the volume \(V\) occupied by the particle,
\begin{equation}
	W_w = w V ,
	\label{EQ:WW:WV}
\end{equation}

with the energy density \(w\) of an electromagnetic wave equal to the square of the electric field \({\bf E}_w\) times electric constant \(\varepsilon_0\).
\begin{equation}
	w = \varepsilon_0 {\bf E}_w^2 = \varepsilon_0 E_w^2
	\label{EQ:W:DENSITY}
\end{equation}

Equations (\ref{EQ:W:DENSITY}) and (\ref{EQ:WW:WV}) plugged in into Eq.\ (\ref{EQ:WW:WG}) yield
\begin{equation}
	\varepsilon_0 E_w^2 \: V = W_g
	\label{EQ:E2:WG}
\end{equation}

The volume \(V\) of the particle, occupying an ellipsoid with diameters \(d_x\) and \(d_0\), is given by Eq.\ (\ref{EQ:VOLUME}).
\begin{eqnarray}
	V &=& \frac{4}{3} \pi \: \frac{d_x}{2} \: \frac{d_0}{2} \: \frac{d_0}{2}
	\nonumber \\ \nonumber \\
	V &=& \frac{1}{6} \pi d_x d_0^2
	\label{EQ:VOLUME}
\end{eqnarray}


With Eq.\ (\ref{EQ:EG:V}) for the energy \(W_g\) of the particle, we obtain Eq.\ (\ref{EQ:E:F0:D0:GAMMA}) for the magnitude \(E_w\) of the electric field of the photons.
\begin{eqnarray}
	\varepsilon_0 E_w^2 \frac{1}{6} \pi d_x d_0^2 & = & \gamma \: 2 h f_0 \nonumber \\
	E_w  & = & 2 \sqrt{\frac{3 \: \gamma \: h f_0}{\varepsilon_0 \: \pi d_x d_0^2}} \label{EQ:E:F0:D0:GAMMA}
\end{eqnarray}

\subsection{Elimination of \(E_w\), \(d_x\), and \(d_0\)}

Now, we can substitute \(E_w\) in Eq.\ (\ref{EQ:GAUSS:5}) by the right-hand side of Eq.\ (\ref{EQ:E:F0:D0:GAMMA}) and obtain
\begin{eqnarray}
	q  &=&   \varepsilon_0 \: C_E \:2\: \sqrt{\frac{3 \: \gamma \: h f_0}{\varepsilon_0
					 \: \pi d_x d_0^2}} \: \frac{d_0}{2} \:  \frac{d_x}{2} \: \pi^2 \nonumber \\
	q  &=&  \: \frac{C_E}{2} \: \sqrt{ \varepsilon_0 \: 3 \: \gamma \: h f_0 d_x \pi^3 }.
			  	\label{EQ:GAUSS:6}
\end{eqnarray}

Substituting \(d_x\) by the right-hand side of Eq.\ (\ref{DX:V}) cancels the frequency \(f_0\), and we obtain expression (\ref{EQ:Q:PARTICLE}) for the charge of the particle, \emph{independent} of its energy.
\begin{equation}
	\boxed{q  = \frac{C_E}{2} \: \sqrt{ 3 \: n_g \pi^3 \varepsilon_0 h c_0}}	\label{EQ:Q:PARTICLE}
\end{equation}

The dimensionless constant \(C_E\) gives the ratio of the mean value \(E_q\) of the charge vector field \({\bf E}_q\) at the surface of the particle to the mean square \(E_w\) of the electric field \({\bf E}_w\) derived from the energy of the particle. \(\varepsilon_0\), \(h\), and \(c_0\) are fundamental constants and the squareroot of their product has, indeed, the units of a charge, so we refer to the expression \( \sqrt{\varepsilon_0 h c_0}\) as the ``Planck charge'' denoted by the symbol \(e_h\), Eq.\ (\ref{EQ:PLANCK})
\begin{equation}
	e_h = \sqrt{\varepsilon_0 h c_0} , \label{EQ:PLANCK}
\end{equation}

\noindent with a numerical value of \(e_h\) = 1.326\;211\,23\(\times 10^{-18}\) As.
The remaining parameters can be summarized to the constant \(C_W\) given by
\begin{equation}
	C_W = \frac{C_E}{2} \: \sqrt{ 3 \: n_g \pi^3} , \label{EQ:CW} \vspace*{5 mm}
\end{equation}

then Eq.\ (\ref{EQ:Q:PARTICLE}) aquires the simple form (\ref{EQ:Q:CWNGEH}).
\begin{equation}
	q  = C_W e_h\label{EQ:Q:CWNGEH}
\end{equation}

To adjust the wave model of charge to the elementary charge, we have to equate \(q\) with \(e\) and resolve Eq.\ (\ref{EQ:Q:CWNGEH}) with respect to constant \(C_W\). Because charges smaller than \(e\) never have been observed, we assume a wave count \(n_g\) of 1 for the electron, thus constant \(C_W\) is simply the ratio of \(e\) and \(e_h\).
\begin{equation}
	C_W = \frac{e}{e_h} = 0.120\:808\:547 \label{EQ:CW:EVAL}
\end{equation}

\section{Conclusions}

We took the idea of captured photons from the wave model and calculated the electric field on the surface of a particle. The net flux of this electric field is equal to the charge of the particle, and our rather simple model showed that the charge is independent from its energy, i.e., independent from its mass, just like observed in nature. But we will not suppress the difficulties with this model.

Generally, the field vector \({\bf E}_w\) of the electromagnetic wave oscillates at the frequency of the captured photons, and subsequently, the charge field vector \({\bf E}_q\) should also oscillate. Then the mean of the field vector, sampled over a sufficient long time, should be zero, thus making the net flux to vanish and no charged particle should be observed. The captured photons must have a property rather different from free running electromagnetic waves. It seems like the captured photons being in a steady state, i.e., the time derivative of the electric field appears to be zero \(\partial {\bf E}_P / \partial t = 0\). Another problem is that electrons, protons, and most of the other particles possess spin, but our model says nothing about the spin of a particle. The future development must solve these problems in order to achieve a model, where presently known properties of particles are adequately represented.

Regarding Eqs.\ (\ref{EQ:FH:V}) and (\ref{EQ:FR:V}) the wave model assumes a preferred frame of reference, and apparently it violates Lorentz invariance. Obviously, the transformation of coordinates applicable here must be different from Lorentz transformation, hence the Maxwell equations become changed and the objection must be taken serious that the wave model probably doesn't comply with some observable phenomena. Recently, it has been demonstrated \cite{Puccini:2003, Buonaura:2004, Weiss:physics/0606242} that a kind of a Galilean transformation, the medium transformation, can be successfully applied on Maxwell equations and the results are compliant with observable phenomena. This medium transformation primarily applies on the wave model, leading to properties of space-time equal to those of Lorentz ether. But there is no discrepancy with relativity, because Lorentz transformation can also be applied on the wave model, it tolerates both transformations.



\bibliography{weiss}
\end{document}